\documentclass[iop]{emulateapj}
\usepackage{amsmath}
\usepackage{subfigure}
\usepackage{graphicx}

\graphicspath{{figs/}}

\newcommand{\vsp}[1]{ \mathbf #1 }

\newcommand{\figext}[1]{#1.eps}

\makeatletter
\def\ScaleIfNeeded{%
  \ifdim\Gin@nat@width>\linewidth
  \linewidth
  \else
  \Gin@nat@width
  \fi
}
\makeatother

\shorttitle{Spectral and Intermittency Properties of Relativistic
  Turbulence}

\shortauthors{J. Zrake and A. I. MacFadyen}

\begin{document}

\title{Spectral and Intermittency Properties of Relativistic
  Turbulence}
\author{Jonathan Zrake and Andrew I. MacFadyen}

\affil{Center for Cosmology and Particle Physics, Physics Department,
  New York University, New York, NY 10003, USA}


\begin{abstract}
  High-resolution numerical simulations are utilized to examine
  isotropic turbulence in a compressible fluid when long wavelength
  velocity fluctuations approach light speed. Spectral analysis
  reveals an inertial sub-range of relativistic motions with a broadly
  5/3 index. The use of generalized Lorentz-covariant structure
  functions based on the four-velocity is proposed. These structure
  functions extend the She-Leveque model for intermittency into the
  relativistic regime.
\end{abstract}

\maketitle

\section{Introduction}
A rapidly accumulating body of astronomical observations indicates the
presence of relativistic turbulence in a variety of astrophysical
systems. Cosmological gamma-ray bursts (GRBs) accelerate relativistic
jets with energies exceeding the kinetic energy of a supernova
($10^{51-52}$ergs) to Lorentz factors $\gamma \gtrsim 100$ and contain
internal fluctuations with $\delta \gamma \sim 2$
\citep{Piran:2004p4605}, while jets from active galactic nuclei and
micro-quasars are found to have $\gamma \gtrsim 10$ and $\gamma
\lesssim 2$ respectively. Recent discoveries indicate that a broad
class of stellar explosions, including those associated with X-ray
transients and low-luminosity GRBs \citep{sn1998bw,grb100316D}, and
even some supernovae lacking high-energy emission
\citep{Soderberg:2010p4890,paragi2010}, also accelerate significant
amounts of relativistic material. These relativistic flows are highly
susceptible to turbulence due to the extremely high inferred Reynolds
numbers and the abundant presence of shear due to observed flow
asymmetries.

In recent models for fluctuating emission from GRBs, the relativistic
turbulence is proposed to be directly observable, as relativistic
eddies beam their radiation into narrow cones which sweep by the
observers' line of sight \citep{Narayan:2009p4019,Lazar:2009p4221}.
In addition, the central engine for short-duration GRBs likely
involves the turbulent merging of neutron stars which are known
sources of gravitational radiation \citep{Hulse:1975p5016}. Knowledge
of the properties of turbulence present in these sources
\citep{Melatos:2010p4963} may aid with detection and interpretation of
gravitational waves from these sources when interferometers such as
advanced LIGO become operable in the near future. Relativistic
turbulence is also likely generated by phase transitions in the early
universe and may be produced in heavy ion colliders and laser
experiments.

\begin{figure*}
  \centering \subfigure{
    \includegraphics[width=3.4in]{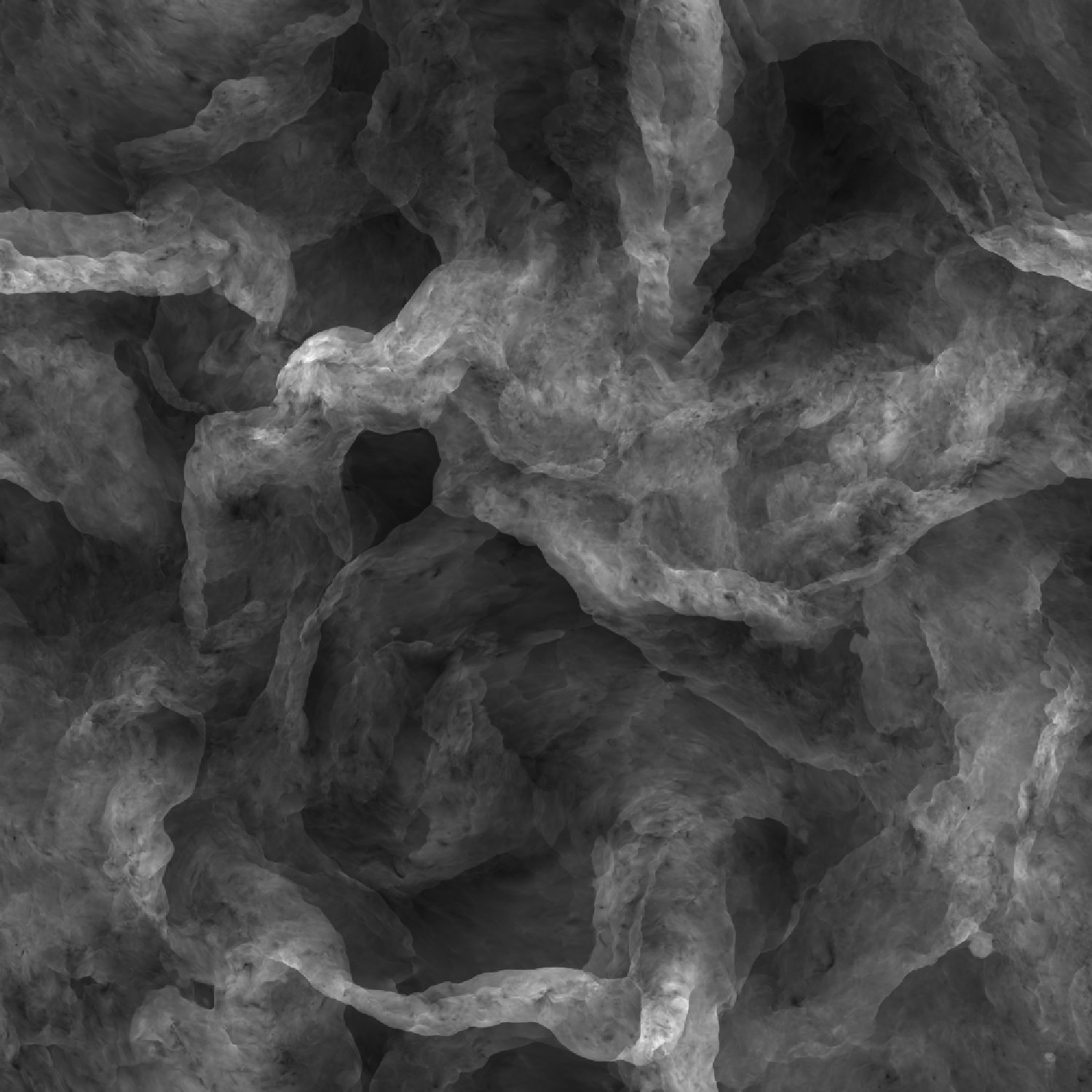}
  }
  \subfigure{
    \includegraphics[width=3.4in]{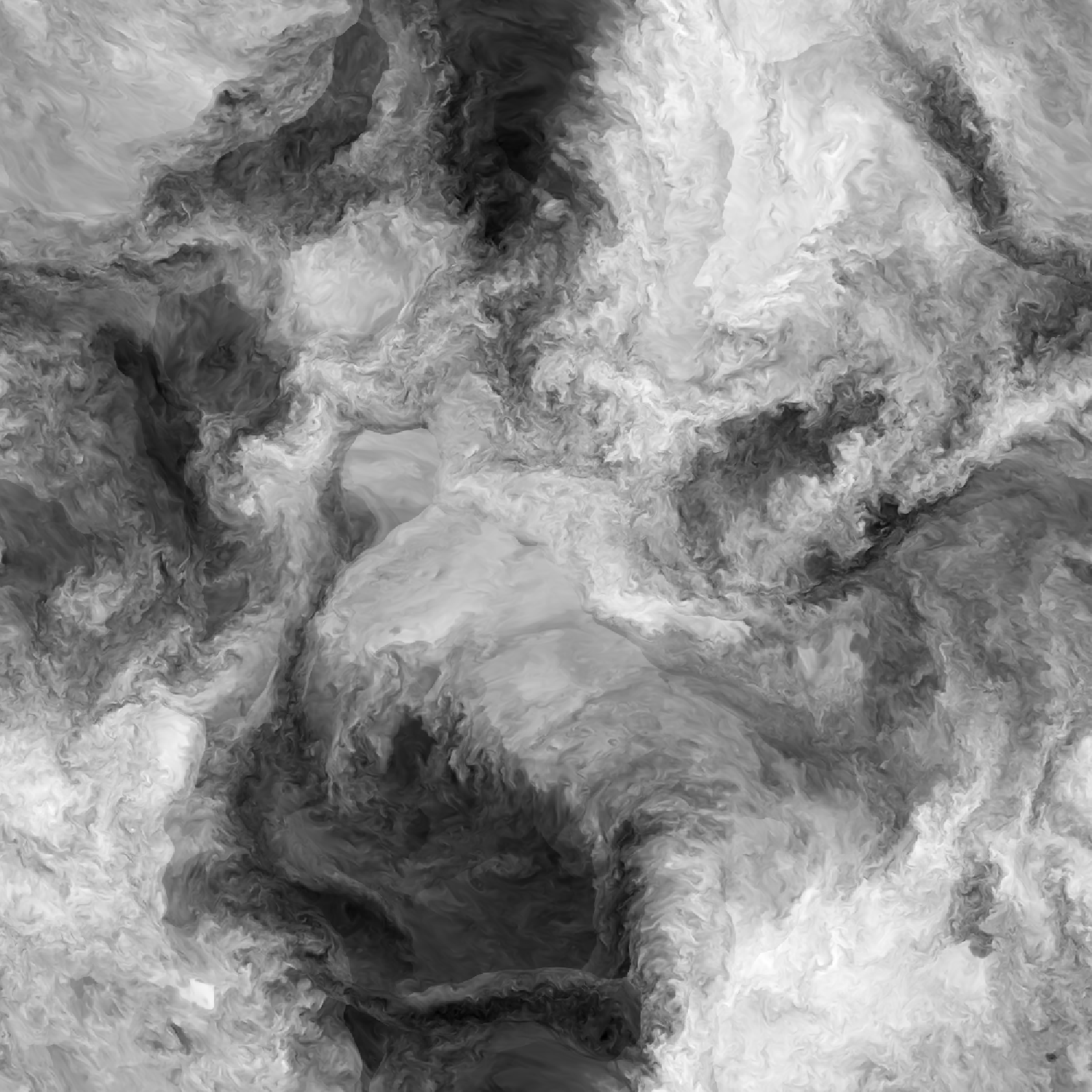}
  }
  \caption{Slices of the co-moving density (left) and four-velocity
    component out of the page (right), indicating the structure of
    isotropic relativistic turbulence at a resolution of
    $2048^3$. Density is scaled to the 1/2 power to elucidate
    structures in the evacuated regions. The grayscale is from minimum
    (darkest) to maximum (brightest).}
  \label{fig:images}
\end{figure*}

The fundamental properties of kinematically relativistic hydrodynamic
turbulence have not, until now, been investigated. Turbulence in fully
relativistic MHD has been studied for decaying
\citep{Zhang:2009p3418,Inoue:2011p5099,Beckwith:2011p4448} and driven
\citep{Zrake:2011p5077,Zrake:2012p5098} flows. Also, simulations
ignoring the ﬂuid rest mass have been done in the magnetized
\citep{Cho:2005p4453} and unmagnetized \citep{Radice:2012p5041}
cases. Fundamental questions relating to the non-linear behavior of
the SRHD system remain to be answered. For example, it has not
previously been known whether relativistic velocity excitations
survive to small scales or are suppressed near the outer scales of a
turbulent flow. If relativistic motions persist deeply into the
cascade, then universal behavior may be expected. However,
relativistic statistical measures are required to characterize its
scaling and intermittency properties. For example, the power spectrum
of three-velocity cannot obey a power law above the scale
$\ell_\gamma$ where the velocity differences asymptotically approach
light speed. Traditional two-point structure functions also break down
at scales $\ell > \ell_\gamma$. Statistical diagnostics built on the
fluid four-velocity, $u = (\gamma c, \gamma \vsp{v})$, are thus
needed.

In this Letter, we present high-resolution simulations of isotropic
relativistic turbulence and introduce Lorentz covariant scaling
diagnostics to analyze its basic properties. Questions to be addressed
include (1) Does relativistic turbulence display universal behavior?
(2) Can its intermittency be described in terms of known
phenomenological models? (3) How deeply into the cascade do velocity
fluctuations remain relativistic?

In order to address these questions, numerical integrations of the
special relativistic hydrodynamics (SRHD) equations were conducted on
the Pleiades supercomputer at the Ames research center on lattice
sizes up to $2048^3$, utilizing more than 2,000,000 CPU-hours. The
SRHD equations are cast in conservation form for the particle number
$\nabla_\nu N^\nu = 0$ and energy-momentum tensor $\nabla_\nu T^{\mu
  \nu} = S^\mu$ and closed by the adiabatic equation of state $p =
\rho e (\Gamma - 1)$ where $\Gamma = 4/3$, $e$ is the specific
internal energy, $\rho$ is the co-moving mass density and $T^{\mu\nu}
= \rho h u^\mu u^\nu + p g^{\mu\nu}$, and $h = c^2 + e + p/\rho$. The
source term $S^\mu=\rho a^\mu - \rho (e/e_0)^4 u^\mu$ includes
injection of energy and momentum at the large scales and the
subtraction of internal energy (with parameter $e_0=0.25$) to permit
stationary evolution. Vortical modes at $k/2\pi \le 3$ are forced by
the four-acceleration field $a^\mu = \frac{du^\mu}{d\tau}$ which
smoothly decorrelates over a large-eddy turnover time, as described in
\cite{Zrake:2012p5098}. The numerical scheme employed is the 5th-order
accurate weighted essentially non-oscillatory (WENO) conservative
finite-difference scheme which has been extensively tested for SRHD
\citep{Zhang:2006p7} and is implemented in the Mara code
\citep{Zrake:2012p5098} with improved smoothness indicators
\citep{Shen:2010p3525}. The simulations take place on the periodic
cube of size $L$, and were evolved at stationarity for at least 5
large eddy turnover times, with the exception of the $2048^3$ run
which was evolved for 1/2 of a turnover time. Each model at resolution
$N^3$ was initialized from stationary turbulence at resolution
$(N/2)^3$, after which a full turnover time is required to
re-establish stationarity. Table \ref{tab:summary} summarizes the
simulation parameters for the $2048^3$ run.

\section{Results}
Images of the fully developed turbulent fields are shown in Fig.
\ref{fig:images}. The co-moving density (left panel) contains
fluctuations above and below the mean by factors of 16. The highest
density structures are sheet-like, but not fully two-dimensional and
trace the front between clouds of fluid colliding with relativistic
velocities. Shocks and contact discontinuities are evident where the
density field changes abruptly. In compressible turbulence, these
shocks indicate the geometry of the most aggressively dissipating
structures \citep{Kritsuk:2007p3858}. These shocks are more intense in
a relativistic gas because their density contrasts may be arbitrarily
large, $\sigma = \frac{\Gamma + 1}{\Gamma - 1} + \frac{\Gamma}{\Gamma
  - 1} e/c^2$ whereas in the non-relativistic limit the density ratio
is bounded by $\frac{\Gamma + 1}{\Gamma - 1}$. The transverse
four-velocity component (right panel) reveals a highly structured
hierarchy of shearing motions down to the smallest scales. It also
exhibits remarkable coherence over some regions, where large clouds of
fluid are in relativistic motion into (or out of) the page. Fluid
elements separated by 1/100, 1/10 and 1/5 of the image are, on
average, in relative motion with four-velocities of 1/2, 1 and 2
respectively.

\begin{deluxetable}{lcc}
\tablecolumns{3}
\small
\tablewidth{0pt}
\tablecaption{Simulation parameters}
\tablehead{\colhead{Parameter} & \colhead{Symbol} & \colhead{Value}}

\tablecomments{Simulation parameters are taken from the $2048^3$
  run. The relativistic Mach number is defined to be
  $\mathcal{M}=(\gamma \beta)_{fluid} / (\gamma
  \beta)_{sound}$. Quantities with over-bars are volume averaged at
  each iteration, and all quantities are time-averaged during
  stationary evolution.}

\startdata
  Box size & $L$ & - \nl
  Driving scale scale & - & $L/2$ \nl
  Relativistic scale & $\ell_\gamma$ & $0.16L$ \nl
  Mean pairwise-relative Lorentz factor & $\bar{\gamma}_{rel}$ & 3.10 \nl
  Mean lab-frame Lorentz factor & $\bar{\gamma}$ & 1.67 \nl
  Max lab-frame Lorentz factor & $\gamma_{max}$ & 5.73 \nl
  Mean relativistic Mach number & $\bar{\mathcal{M}}$ & 2.67 \nl
  Max relativistic Mach number & $\mathcal{M}_{max}$ & 22.9 \nl
  Mean pressure to mean density ratio & $\bar{p} / \bar{\rho} c^2$ & 0.41 \nl
  Lower 1\% density quantile & - & $0.027 \bar{\rho}$ \nl
  Upper 1\% density quantile & - & $4.89 \bar{\rho}$ \nl
\enddata
\label{tab:summary}
\end{deluxetable}

Turbulent flows at sufficiently large Reynolds number are known to
contain a range of scales over which the behavior is universal,
i.e. statistically independent of energy injection and dissipation
effects at the largest and smallest scales respectively. This range is
referred to as the inertial sub-interval, and is often identified by
self-similar behavior in the power spectrum $P(k)$ of the velocity
field. The 1941 theory of Kolmogorov (K41) \citep{Kolmogorov:1941p912}
for incompressible, non-relativistic turbulence predicts that $P(k)
\propto k^{-5/3}$ in the inertial interval. In relativistic flows the
amplitude of velocity fluctuations cannot exceed $c$, so the power
spectrum of velocity cannot obey a power law to arbitrarily large
scales. However, the physics of a relativistic cascade may still be
universal, and it could be revealed by an appropriate choice of
generalized diagnostics, such as the power spectrum of fluid
four-velocity.

\begin{figure*}
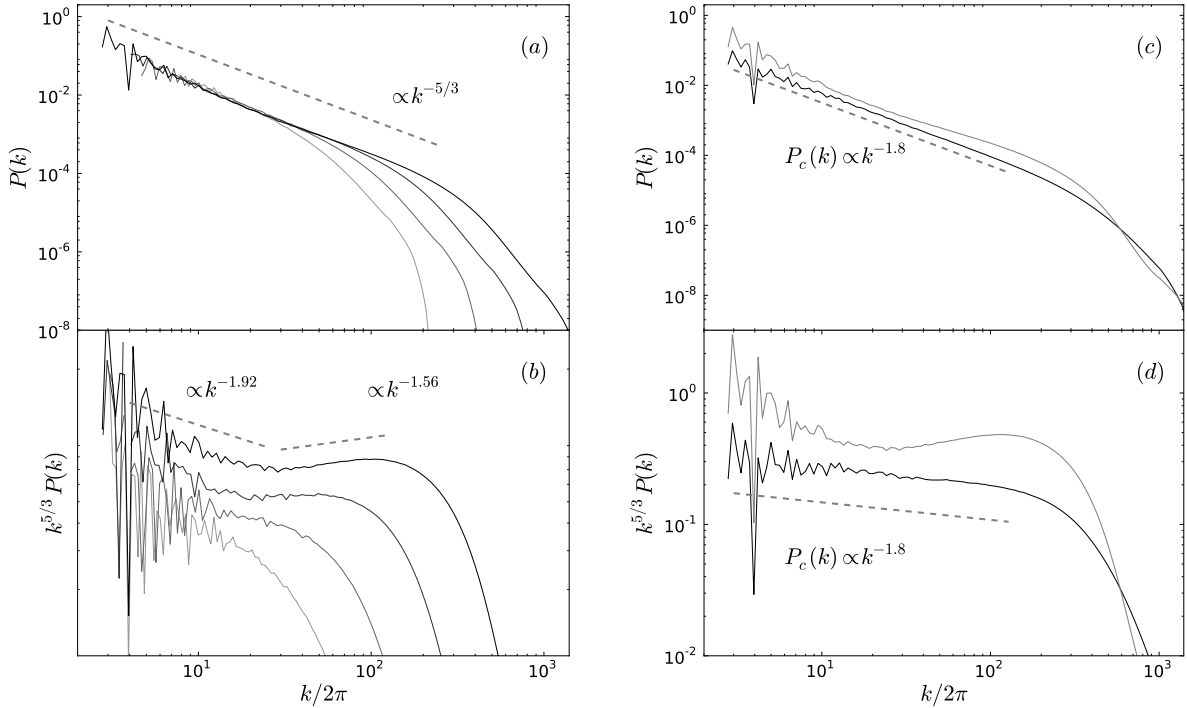

  \centering

  \subfigure{
    \includegraphics[width=3.1in]{\figext{fig1}}
  }
  \subfigure{
    \includegraphics[width=3.1in]{\figext{fig2}}
  }
  \caption{$(a)$ Power spectrum of the four-velocity field at
    resolutions $256^3$, $512^3$, $1024^3$, and $2048^3$ (increasing
    in weight from gray to black). $(b)$ The same data as in $(a)$ but
    compensated by $k^{5/3}$ and offset, stretched to exaggerate
    deviations from a 5/3 law. An arbitrary vertical offset is given
    to each curve in order to clarify the spectral shape between
    $k/2\pi (=1/L) = 10$ and $100$. $(c)$ Power spectrum of the
    Helmholtz decomposed four-velocity field. The compressive
    component $P_c(k)$ (black) follows a power law with index
    $1.80$. $(d)$ The same data as in $(c)$ but compensated by
    $k^{5/3}$.}
  \label{fig:pspecs}
\end{figure*}


Figs. \ref{fig:pspecs}a and \ref{fig:pspecs}b show the power spectrum
$P(k)$ of four-velocity at lattice sizes of $256^3$, $512^3$,
$1024^3$, and $2048^3$, where
\begin{equation}
  P(k)dk = \sum_{\vsp{k} \in dk}{\tilde{\vsp{u}}_{\vsp{k}} \cdot
    \tilde{\vsp{u}}^*_{\vsp{k}}}
\end{equation}
is computed from the discrete Fourier transform
\begin{equation}
  \tilde{u}^\mu_{\vsp{k}} =
  \frac{1}{N^3}\sum_{l,m,n}{u^\mu(\vsp{x}_{l,m,n})e^{-i
      \vsp{k}\cdot \vsp{x}_{l,m,n}}}
\end{equation}
of the four-velocity field. Throughout the text, boldface is used to
denote the spatial components of a four-vector. The normalization by
$N^3$, where $N$ is the number of lattice points per direction,
guarantees that $P(k)$ satisfies $\int{P(k) dk} = \langle \vsp{u}
\cdot \vsp{u} \rangle$. We find evidence for an inertial interval of
relativistic velocity fluctuations between 1/10 and 1/100 of the
largest scale. As the resolution increases from $256^3$ to $1024^3$, a
short interval obeying the 5/3 law emerges. But the subsequent
resolution of $2048^3$ reveals a more featureful spectrum of
four-velocity, consisting of a broken power-law which is steeper
(1.92) at the largest scales and shallower (1.56) at moderate
scales. This may be due to the bottleneck effect which is
characterized by an accumulation of power at the small-scale end of
the inertial range \citep{Falkovich:1994p3874}. There has been
substantial effort to distinguish true inertial scaling from the
contamination of bottlenecks \citep{Beresnyak:2009p4471}. We report
that the power spectrum of four-velocity is \emph{broadly} 5/3, owing
to the fact that each higher resolution adds additional scales which,
on the average, obey 5/3 scaling. In relativistic turbulence,
asymptotically converged scaling behavior may not yield a single power
law. This is because a relativistic cascade contains a new
dimensionless number, the Lorentz factor, at each scale. As a simple
illustration, a cursory application of the K41 dimensional argument to
relativistic eddies reveals that the Lorentz factor at a scale $\ell$
satisfies
\begin{equation}
  \frac{(\gamma_\ell - 1)^3(\gamma_\ell + 1)}{\gamma_\ell^2} =
  C \epsilon^2 \ell^2 / c^6.
\end{equation}
where $\epsilon$ is the energy injection rate per unit mass and $C$ is
analogous to the Kolmogorov constant. This relation does not admit a
single power law solution. Instead, solutions are $\gamma_\ell \propto
\ell$ for large $\gamma$, and $\gamma_\ell \propto \ell^{2/3}$ for
small $\gamma$.

Unlike the total fluid four-velocity, the compressive four-velocity
component $u^\mu_c$ is strictly self-similar over the inertial
interval. This is evidence that a local cascade of acoustic modes is
operating across those scales. The Helmholtz decomposition of spatial
components of four-velocity is done in the Fourier domain, with the
compressive part $\tilde{\vsp{u}}^c_{\vsp{k}} =
\tilde{\vsp{u}}_{\vsp{k}} \cdot \vsp{k}/k$ and the solenoidal part
$\tilde{\vsp{u}}^s_{\vsp{k}} = \tilde{\vsp{u}}_{\vsp{k}} -
\tilde{\vsp{u}}^c_{\vsp{k}}$. As shown in Figs. \ref{fig:pspecs}c and
\ref{fig:pspecs}d, the power-law range of dilatational power spectrum
$P_c(k)$ extends through the energy injection scale. This is because
only the vortical modes are being directly excited. Its slope is 1.80,
which is in contrast with the supersonic limit of highly compressible
turbulence, where all the compressive power is in shocks, and the
dilatational power spectrum $P_c(k) \propto k^{-2}$. The power
spectrum of solenoidal $P_s(k)$ four-velocity is not at all
self-similar, and accounts for the majority of power ($82\%$)
throughout the cascade. We remark that thermally relativistic
turbulence admits a novel mechanism for energy exchange between the
vortical and compressive motions of the fluid. The mechanical work
done in compressing a fluid volume goes into internal energy, which
for a relativistic gas enhances its inertia through the coupling of
total enthalpy to fluid four-velocity expressed by the term $\rho h
u^\mu u^\nu$ in the energy-momentum tensor. This additional mechanism
of energy transfer adds qualitatively new dynamics to the turbulent
cascade. In particular, it may break the statistical decoupling
between the compressive and vortical cascades that was recently
discovered for non-relativistic compressible turbulence
\citep{Aluie:2011p4489}. The signature of this effect is proposed to
be an enhanced coherency among the phases of similar-scale shearing
and dilating velocity modes, that can be determined from the
bi-spectrum of fluid four-velocity.

\begin{figure}
  \centering
  \includegraphics[width=3.4in]{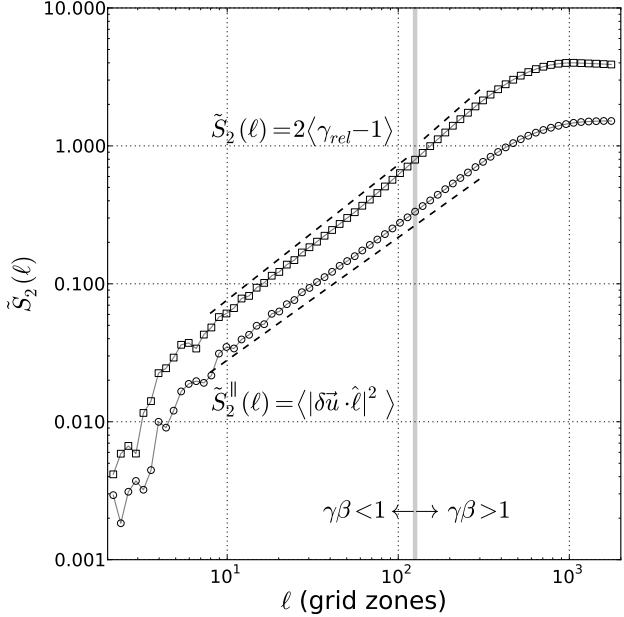}
  \caption{The generalized structure functions of four-velocity
    $\tilde{S}_2^\parallel(\ell)$ (longitudinal projection, squares) and
    $\tilde{S}_2(\ell)$ (circles) as a function of the separation $\ell$ for
    resolution $2048^3$. The vertical gray bar marks the relativistic scale
    $\ell_\gamma = 0.06L$, below which velocity fluctuations become
    sub-relativistic.}
  \label{fig:structure-functions}
\end{figure}
\begin{figure}
  \centering
  \includegraphics[width=3.4in]{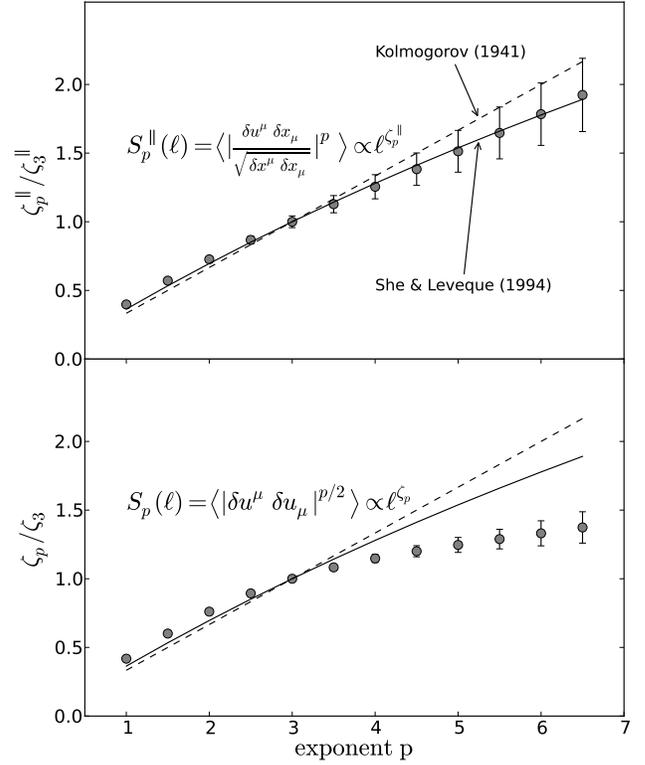}
  \caption{Dependence of $\zeta_p$ on the exponent $p$ for the
    power-law scaling of longitudinal $\tilde{S}^\parallel_p$ (top)
    and total $\tilde{S}_p(\ell)$ (bottom) structure
    functions. $\zeta_p$ is obtained by a least-squares fit of the
    structure functions between $\ell=1/10$ and 1/100 the domain. All
    the scaling exponents are normalized by $\zeta_3$. Error bars
    indicate the standard deviation over 6 equally-spaced stationary
    snapshots at $1024^3$. The intermittency models of K41 and SL94
    are shown in dashed and solid lines respectively.}
  \label{fig:she-leveque}
\end{figure}

Modern descriptions of turbulence have recognized that two-point
correlations of velocity are not normally distributed around the mean
fluctuating fluid velocity
\citep{Kolmogorov:1962p4545,Kraichnan:1990p5021,Cao:1996p5030}. This
effect is referred to as intermittency, and represents an important
departure from the K41 theory. Intermittency may be quantified by
examining higher moments of the probability distribution function
(PDF) of the velocity increments at various scales $\ell$. In
particular, the 1994 model of She-Leveque (SL94) \citep{She:1994p4537}
has had great success in predicting the scaling exponents $\zeta_p$
for the velocity structure functions $S_p( \ell)$ of order $p$. We
have found that relativistic turbulence also displays intermittency,
and that the SL94 model provides an excellent description, provided
that velocity increments are properly defined to accommodate
relativistic fluctuations. We generalize the total velocity structure
function as
\begin{equation}\label{eqn:Spar}
  \tilde{S}_p(\ell) = \langle |\delta u^\mu \delta u_\mu|^{p/2}
  \rangle
\end{equation}
where $\delta u^\mu = u_2^\mu - u_1^\mu$ and $\ell = (\delta x^\mu
\delta x_\mu)^{1/2}$. Four-velocity pairs are chosen to be
simultaneous in the global center-of-momentum frame so that $\delta x^0 =
0$. Note that $\tilde{S}_p(\ell) = \langle |2(\gamma_{rel} - 1)|^{p/2}
\rangle$ where $\gamma_{rel}$ is the Lorentz factor of one fluid
element as measured in the rest frame of the
other. $\tilde{S}_p(\ell)$ becomes $\langle |\vsp{v}_2 - \vsp{v}_1|^p
\rangle$ in the non-relativistic limit. We also generalize the
longitudinal structure function
\begin{equation}\label{eqn:Stot}
  \tilde{S}_p^\parallel(\ell) = \langle |\frac{\delta u^\mu \delta
    x_\mu }{(\delta x^\mu \delta x_\mu)^{1/2}}|^p \rangle
\end{equation}
which can also be written as $\langle |\delta \vsp{u} \cdot
\hat{\ell}|^p \rangle$, where $\hat{\ell} = \boldsymbol{\ell}/\ell$,
because of pair simultaneity.

We utilize $\tilde{S}_2(\ell)$ to determine the scale $\ell_\gamma$ to
which relativistic velocity fluctuations persist. The kinematic effects
of relativity are important when the \emph{relative} four-velocity
$\gamma \beta$ between fluid elements exceeds 1, i.e. $\langle
\gamma_{rel} \rangle > \sqrt{2}$ (or $\tilde{S}_2(\ell) = 2\sqrt{2} -
2 \approx 0.83$). As shown in Fig. \ref{fig:structure-functions}, this
occurs at $\ell_\gamma = 0.06$, or 1/16 of the outermost
scale. $\tilde{S}_2(\ell)$ shows two regimes of power-law scaling,
having exponents of $\zeta_2 = 0.98$ below $\ell_\gamma$ and $\zeta_2
= 1.09$ above $\ell_\gamma$. The longitudinal structure function
$\tilde{S}_2^\parallel(\ell)$ obeys a power law with index
$\zeta^\parallel_2 = 0.89$ between 1/10 and 1/100 of the domain. This
is steep relative to the K41 theory for incompressible
non-relativistic turbulence that predicts $\zeta_2 = 2/3$, but is in
close agreement with what was found in simulations of highly
compressible non-relativistic turbulence \citep{Kritsuk:2007p3858}.

We find that the SL94 intermittency model extends to relativistic
turbulence when the structure function is given by Eqn.
\ref{eqn:Spar}. Fig. \ref{fig:she-leveque} shows that the normalized
exponents $\zeta^\parallel_p/\zeta^\parallel_3$ through $p=6.5$ are
well described by the SL94 relation $\zeta_p = p/9 + 2 -
2(2/3)^{p/3}$. Error bars are obtained from the standard deviation
over 6 snapshots at $1024^3$. The total structure function given in
Eqn. \ref{eqn:Stot}, which measures the statistics of pairwise
relative Lorentz factor, exhibits a far greater degree of
intermittency. We propose that a new intermittency model for the
statistics of relative four-velocities is required, noting that these
statistics are inherently non-Gaussian from consideration of
relativistic kinematics alone. That is to say, a Gaussian random
velocity field does not yield a Gaussian PDF of $\gamma_{rel}$, owing
to the non-additive nature of Lorentz boosts. A subsequent
investigation will examine the shape of the $\tilde{S}_p(\ell)$ PDF at
fixed scales in order to more specifically characterize the source of
the intermittency seen in Fig. \ref{fig:she-leveque}.

\section{Discussion}
In this Letter we have described a numerical investigation of
isotropic turbulence in a kinematically and thermally relativistic
fluid. It is found that relativistic fluctuations persist to 1/16 of
the outermost scale. Spectral analysis reveals an inertial sub-range
of relativistic velocity fluctuations with a broadly 5/3 index. The
compressive component of the four-velocity field obeys power-law
scaling with index 1.80 over nearly two decades. The SL94
phenomenology successfully describes intermittency in relativistic
turbulence, provided the correct definition for the velocity
increments is used.

In closing, we remark that for sufficiently large Reynolds number,
$\ell_\gamma$ is larger than the dissipation scale $\eta$ so that
kinetic energy must be transmitted through a sub-relativistic cascade
before being thermalized. In this case the ratio $\ell_\gamma / L$ may
be independent of the dissipation mechanism, and should thus be
dictated solely by the Lorentz factor of the energy bearing
eddies. The exact functional form of this dependence could be
constrained by a parameter study over the outer scale Lorentz
factor. Doing so would yield insight into the nature of astrophysical
turbulent cascades operating far into the relativistic regime.

\acknowledgments

This research was supported in part by the NSF through grant
AST-1009863 and by NASA through grant NNX10AF62G issued through the
Astrophysics Theory Program. Resources supporting this work were
provided by the NASA High-End Computing (HEC) Program through the NASA
Advanced Supercomputing (NAS) Division at Ames Research Center. We
thank Andrei Gruzinov and Paul Duffell for insightful discussions.


\end{document}